\documentclass[12pt,a4paper]{article}
\usepackage{epsfig}
\usepackage{amstex}
\usepackage{amssymb}

\def\as{\alpha_s}
\def\gl{\tilde{g}}
\def\sq{\tilde{q}}
\def\st{\tilde{t}}
\def\sqb{\bar{\tilde{q}}}
\def\ms{m_{\tilde q}}
\def\mg{m_{\tilde g}}

\begin{document}
\thispagestyle{empty}
\hfill\vbox{\hbox{hep-ph/9611232}
            \hbox{November 1996}}

\setcounter{footnote}{0}

\vspace*{1cm}
\begin{center}
  \textbf{\Large \sc PROSPINO}\\[0.5cm]
  {\large \sc a program for the \\[0.2cm]
    {\bf PRO}duction of {\bf S}upersymmetric {\bf P}articles\\[0.2cm]
    {\bf I}n {\bf N}ext-to-leading {\bf O}rder qcd}\\[4cm]
\def\thefootnote{\fnsymbol{footnote}}%
  {\sc W.~Beenakker}\footnote{Research supported by a fellowship of the 
                              Royal Dutch Academy of Arts and Sciences.}%
\def\thefootnote{\arabic{footnote}}%
\setcounter{footnote}{0}%
\footnote{wimb@@lorentz.leidenuniv.nl}, 
  {\sc R.~H\"opker}\footnote{hoepker@@x4u2.desy.de}
  {\sc and M.~Spira}\footnote{spira@@cern.ch} \\
\vspace{0.5in}
$^1$ \emph{Instituut--Lorentz, P.O.~Box 9506, NL--2300 RA Leiden, 
           The Netherlands}\\
$^2$ \emph{Deutsches Elektronen-Synchrotron DESY, D--22603 Hamburg,
           Germany}\\ 
$^3$ \emph{CERN, TH Division, CH--1211 Geneva 23, Switzerland}\\[4cm] 

\textbf{Abstract}
\end{center}

A Fortran-program for the production cross-sections of squarks and
gluinos at hadron colliders is described. It includes full
next-to-leading order SUSY-QCD corrections to all possible final
states ($\sq\sqb, \gl\gl, \sq\gl, \sq\sq$). The program allows to
calculate total cross-sections as well as differential distributions
in the transverse momentum $p_t$ and the rapidity $y$ of one of the
outgoing particles. In addition cuts in $p_t$ and $y$ can easily be
implemented.

\pagebreak
\setcounter{footnote}{0}

\section{Introduction}
One of the important quests of high-energy physics is the search for
supersymmetric particles. Squarks and gluinos, the supersymmetric partners
of the quarks and gluons, can be produced at
present and future hadron colliders, if their masses are in the
accessible range. The next-to-leading order (NLO) strong (SUSY-QCD)
corrections increase the cross-sections for
the various production processes by up
to a factor of two with respect to the leading-order (LO) predictions and 
reduce the dependence on the renormalization and 
factorization scale significantly \cite{BHSZ1,BHSZ2}. Therefore a theoretically
stable prediction for the cross-sections has to include these
corrections.

\smallskip %
In this paper we present the Fortran-program package PROSPINO, which 
calculates the production cross-sections for squarks ($\sq$) and gluinos 
($\gl$) in LO and NLO. This program complements the
event-generator ISAJET \cite{BPPT}, which includes the complete
production and decay processes for supersymmetric particles in Born
approximation. The calculation has been performed for the reactions
\begin{equation}
  pp/p\bar{p} \to\sq\sqb, \gl\gl, \sq\gl, \sq\sq \qquad \qquad (\sq\neq\st)
  \label{pro}
\end{equation}
in a supergravity-inspired model in which all squarks have one
common mass\footnote{This assumption is not correct for the stops.
  Nevertheless the induced error is very small, because we are
  including stops only in internal loops and not as final-state particles.
  The calculations for stop production are in progress~\cite{BHPZ} and will
  be implemented in PROSPINO in due course.}.
The program calculates differential cross-sections in the transverse
momentum $p_t$ and the rapidity $y$, as well as total cross-sections
with possible cuts in $p_t$ and $y$. The masses of the squarks
($\ms$), gluinos ($\mg$) and top quarks ($m_t$), the 
renormalization/factorization scale ($Q$), the collider type ($pp/p\bar{p}$),
and the set of parton densities can be chosen. For numerical integration the
VEGAS-routine \cite{Lepage} is used.

The program is written in Fortran, and has been tested on several
Unix-workstations. 

\section{The Calculation}
The total hadronic cross-sections for the various final states are
calculated in the following way:
\begin{eqnarray}
  \sigma(pp/p\bar{p} \to \sq + \sqb +X) & = &
  \int_{p_t^{min}}^{p_t^{max}} dp_t \int_{y^{min}}^{y^{max}} dy \,\,
  \frac{d^2\sigma(\sq+\sqb)}{dp_t \,dy} \label{sb}\\
  \sigma(pp/p\bar{p} \to \gl + \gl +X) & = &
  \int_{p_t^{min}}^{p_t^{max}} dp_t \int_{y^{min}}^{y^{max}} dy \,\,
  \frac{d^2\sigma(\gl+\gl)}{dp_t \,dy} \label{gg}\\ 
  \sigma(pp/p\bar{p} \to \sq + \gl +X) & = &
  \int_{p_t^{min}}^{p_t^{max}} dp_t \int_{y^{min}}^{y^{max}} dy \,\,
  \frac{d^2\sigma(\sq+\gl)}{dp_t \,dy} \label{sg}\\ 
  \sigma(pp/p\bar{p} \to \sq + \sq +X) & = &
  \int_{p_t^{min}}^{p_t^{max}} dp_t \int_{y^{min}}^{y^{max}} dy \,\,
  \frac{d^2\sigma(\sq+\sq)}{dp_t \,dy} \label{ss}
\end{eqnarray}
The quantities $p_t$ and $y$ are defined as the transverse momentum
and the rapidity of the \emph{second} particle\footnote{The distributions with
  respect to the \emph{first} particle are identical for squark- and 
  gluino-pair production. For squark--antisquark final states, the program 
  allows to choose the particle for which $p_t$ and $y$ are defined 
  (squark/antisquark/average of both). This facilitates adding the 
  distributions with respect to both final-state particles if squarks and
  antisquarks are not discriminated in the experimental analyses.} 
[e.g.~of the gluino in Eq.~(\ref{sg})].  For the processes
in Eqs.~(\ref{sb}), (\ref{sg}), and (\ref{ss}) the summation over all
possible squark flavors (except stop) and over both squark chiralities
is implicitly understood. In addition Eqs.~(\ref{sg}) and (\ref{ss})
include the sum over the charge-conjugated final states.  The rapidity
distributions are defined by adding the contributions of positive and
negative rapidity.  Thus the integration over $y$ has to include only
the range of positive rapidity [\emph{i.e.}~$y^{min} \ge 0$]. The
factor $1/2$ for identical particles in the final state is included in
the double-differential cross-sections.

The double-differential cross-section can be written as
\begin{equation}
  \frac{d^2\sigma}{dp_t \,dy} = 2 p_t S \sum_{i,j=g,q,\bar{q}}
  \int_{x_1^-}^1 dx_1 \int_{x_2^-}^1 dx_2 \, x_1 f_{i}^{h_1}
  (x_1,Q^2)\, x_2 f_{j}^{h_2} (x_2,Q^2)
  \,\frac{d^2\hat{\sigma}_{ij}(x_1 x_2 S,Q^2)}{dt\, du}
  \label{d2sig}
\end{equation}
where $Q$ is the renormalization/factorization scale\footnote{In NLO
  the scale must be fixed by an external (hadronic) scale. The natural
  value is given by the average mass of the produced massive particles 
  or the transverse mass of the detected final-state particle.}, $i$ and $j$ 
indicate the initial-state partons,
$f_i$ are the parton densities (in the $\overline{MS}$ factorization
scheme), and $\sqrt{S}$ is the center-of-mass energy of the collider.
The invariants $t$ and $u$ are the usual Mandelstam variables, related to the
momentum transfer from the initial-state partons to the detected final-state 
particle~\cite{BHSZ2}. 
The program calculates the double-differential cross-sections in 
Eq.~(\ref{d2sig}) in LO and NLO. The NLO results include the sum of
leading and next-to-leading order contributions. For LO and NLO
various sets of parton densities can be used. As two standard
parametrizations the program includes the GRV94 \cite{GRV94} and the
MRS(A') \cite{MRSA} parton densities. Moreover, the PDFLIB library
\cite{PDF} can be linked to the program, therefore other parton
densities can be included rather easily.

The technical details, like the correct phase-space boundaries or the
subtraction procedure for on-shell intermediate states, can be found
in Ref.~\cite{BHSZ2}.

\section{Computer Implementation}
\subsection{Organization}
The program package PROSPINO consists of several files. Compiling
and linking is done most efficiently
with a \textbf{makefile}\footnote{ The command syntax is: \textbf{make
    'executable'}.} under the operating system Unix. The executable files are:
\begin{description}
\item[totalsb.f] calculation of the cross-sections for
  squark--antisquark production
\item[totalgg.f] calculation of the cross-sections for
  gluino--gluino production
\item[totalsg.f] calculation of the cross-sections for
  squark--gluino production
\item[totalss.f] calculation of the cross-sections for
  squark--squark production  
\end{description}
The files that should be compiled to object files are the following (this is
automatically done by the command \textbf{make library}):
\begin{description}
\item[matrixsb.f] squared matrix elements for squark--antisquark production
\item[matrixgg.f] squared matrix elements for gluino--gluino production
\item[matrixsg.f] squared matrix elements for squark--gluino production
\item[matrixss.f] squared matrix elements for squark--squark production
\item[hadronsb.f] definition of the cross-sections for
  squark--antisquark production 
\item[hadrongg.f] definition of the cross-sections for
  gluino--gluino production 
\item[hadronsg.f] definition of the cross-sections for
  squark--gluino production 
\item[hadronss.f] definition of the cross-sections for
  squark--squark production 
\item[layoutsb.f] print routines for squark--antisquark production
\item[layoutgg.f] print routines for gluino--gluino production
\item[layoutsg.f] print routines for squark--gluino production
\item[layoutss.f] print routines for squark--squark production
\item[integral.f] scalar and angular integrals and the VEGAS
  routine
\item[pdfgrv94.f] GRV94 parton densities\footnote{For charm and bottom
    distributions, the older GRV densities are used.}
\item[pdfmrsap.f] MRS(A') parton densities
\item[initpdf.f] choice of the parton densities (PDFLIB) and $\as$
  in LO and NLO
\end{description}
We strongly recommend to make \emph{cautious} changes only in the
\textbf{total*.f} and \textbf{layout*.f} files.  For the long files
(\textbf{matrix*.f}) some compilers need an increased table size; this
can be changed in the \textbf{makefile} with the parameter longopt.

\subsection{Input Parameters}
In this section the user-setable physical and numerical input
parameters are described.

\begin{description}
\item[MS, MG, MT] ({\it REAL*8}): masses (in GeV) of the squarks (MS), the
                 gluino (MG), and the top quark (MT).
\item[ICOLL] ({\it INTEGER}): type of the hadron collider
  \begin{itemize}
  \item[{\bf 0}] $p\bar p$ collider
  \item[{\bf 1}] $pp$ collider
  \end{itemize}
\item[ENERGY] ({\it REAL*8}): center-of-mass energy of the collider (in GeV).
\item[IPDFSET] ({\it INTEGER}): set of parton densities and the corresponding 
                 $\as$
  \begin{itemize}
  \item[{\bf 0}] GRV94 parton densities 
  \item[{\bf 1}] MRS(A') parton densities 
  \item[{\bf 2}] parton densities from the PDFLIB library\footnote{In
      addition, three things must be arranged: the correct cernlib
      directory must be given in the \textbf{makefile}; the three
      (dummy) routines (PDFSET, PFTOPDG, and ALPHAS2) in
      \textbf{initpdf.f} must be commented out; in the subroutines
      INILO and ININLO in \textbf{initpdf.f} the desired PDFLIB
      parameters must be chosen.}
  \end{itemize}
\item[IFLAVOR] ({\it INTEGER}): initial states
  \begin{itemize}
  \item[{\bf 0}] sum over all initial states
  \item[{\bf 1}] gluon--gluon 
  \item[{\bf 2}] quark--quark (= the sum over $q\bar{q}$, $q'\bar{q}$, $qq$,
                 $q'q$, $\bar{q}\bar{q}$, and $\bar{q}'\bar{q}$)
  \item[{\bf 3}] quark--gluon (= the sum over $gq$ and $g\bar{q}$)
  \item[{\bf 4}] the major contributions\footnote{The initial states that  
                 exist already in LO contribute nearly $100\%$ of the NLO
                 cross-section.  Only these contributions will be calculated
                 here.}
  \item[{\bf 5}] the minor contributions\footnote{The initial states that 
                 start to contribute at NLO yield a tiny contribution to the 
                 NLO cross-sections.  Due to the subtraction procedure for 
                 on-shell intermediate states in some regions of the parameter 
                 space, these contributions may require a higher integration 
                 precision to produce stable results. With this option they can
                 be calculated separately and added to the main contributions.}
  \end{itemize}
\item[ITOTAL] ({\it INTEGER}): the way of calculating the cross-section 
  \begin{itemize}
  \item[{\bf 0}] total cross-section with cuts or differential cross-section 
                 (slower)
  \item[{\bf 1}] only total cross-section without cuts (faster)
  \end{itemize}
\item[ISCAPT] ({\it INTEGER}): type of the default scale
  \begin{itemize}
  \item[{\bf 0}] the average mass of the outgoing massive particles;
                 default for ITOTAL = 1
  \item[{\bf 1}] the transverse mass $\sqrt{m^2+p_t^2}$ of the detected 
                 particle with mass $m$ and transverse momentum $p_t$; 
                 only for ITOTAL = 0 
  \end{itemize}
\item[SCAFAC] ({\it REAL*8}): factor by which the default scale is multiplied 
\item[PTMIN, PTMAX] ({\it REAL*8}): lower and upper bounds of the 
                 transverse-momentum ($p_t$) integration. Only positive values
                 for PTMIN and PTMAX are allowed.
                 The choice PTMIN = PTMAX yields $d\sigma/dp_t$.
\item[YMIN, YMAX] ({\it REAL*8}): lower and upper bounds of the rapidity
                 ($y$) integration. The rapidity distributions are defined by 
                 adding the contributions of positive and negative rapidity, 
                 hence $\mathrm{YMIN}\ge 0$ is required. 
                 The choice YMIN = YMAX yields $d\sigma/dy$.
\item[ICHARCONJ] ({\it INTEGER}): only for squark--antisquark final
  states. The particle for which the differential cross-sections are
  defined 
  \begin{itemize}
  \item[\boldmath{$-1$}] antisquark
  \item[\boldmath{$+1$}] squark  
  \item[{\bf 0}] the average of squark and antisquark 
  \end{itemize}
\item[IONLYLO] ({\it INTEGER}): for calculating only the LO results 
  (\textbf{1}) or both LO and NLO results (\textbf{0}).
\item[ILO, INLO] ({\it INTEGER}): number of VEGAS calls. VEGAS and VEGAS1 will
                 be called by the routine INTEG (VEGAS1 with five times as 
                 many calls as VEGAS).
\item[IPRINT] ({\it INTEGER}): print intermediate results of VEGAS
  iterations (\textbf{10}) or not (\textbf{0}).
\end{description}

\subsection{Results}
The program prints the relevant physical parameters and the hadronic
cross-sections in LO and NLO (in pb). If the lower and upper bounds in $p_t$
and/or $y$ are identical, the results correspond to the
differential cross-sections $d\sigma/dp_t$, $d^2\sigma/dp_t/dy$, or 
$d\sigma/dy$, respectively.

\subsection{How to get the programs}
The program package PROSPINO is available upon request from the authors
or can be picked up from the WWW address:
http://wwwcn.cern.ch/\~{}mspira/. For any comments, questions, or problems 
please contact the authors.

\section{Appendix}
\subsection{Sample File \textbf{totalsb.f}}
\begin{verbatim}
C**********************************************************************
C***                                                                ***
C***  THIS PROGRAM CALCULATES                                       ***
C***  THE TOTAL CROSS-SECTION AND DISTRIBUTIONS                     ***
C***  FOR SQUARK-ANTISQUARK PRODUCTION AT HADRON COLLIDERS          ***
C***  INCLUDING FULL SUSY-QCD CORRECTIONS                           ***
C***                                                                ***
C***  WRITTEN BY: W. BEENAKKER, R. HOPKER AND M. SPIRA              ***
C***                                                                ***
C**********************************************************************

      PROGRAM TOTALSB
      IMPLICIT REAL*8 (A-H,M-Z)
      IMPLICIT INTEGER (I,J)
      COMMON/IOUT/IPRINT
      COMMON/CONST1/S,ENERGY,ALPHAS,MS,MG,MT
      COMMON/CONST2/SCALE,SCAFAC,ICOLL,ISCAPT
      COMMON/CONST3/IPDFSET
      COMMON/CONST5/ILO,INLO,IONLYLO
      COMMON/CUT1/PTMIN,PTMAX
      COMMON/CUT2/YMIN,YMAX
      COMMON/FLAVOR/IFLAVOR,ITOTAL
      COMMON/CHARCONJ/ICHARCONJ

C ---------------------------------------------------------------
C --- INPUT PARAMETERS, CAN BE CHANGED  -------------------------
C ---------------------------------------------------------------

C***  THE MASSES (IN GEV)

      MS = 280.D0      
      MG = 200.D0
      MT = 175.D0

C***  THE COLLIDER TYPE ( P PBAR = 0, P P = 1 )
      
      ICOLL = 0

C***  THE CENTER OF MASS ENERGY (IN GEV)

      ENERGY = 1800D0

C***  THE SET OF PARTON DENSITIES (GRV = 0, MRSAP = 1, PDFLIB = 2)
C***  FOR PDFLIB PLEASE MAKE CHANGES IN SUBROUTINES INILO AND ININLO
C***  IN FILE INITPDF.F

      IPDFSET = 0

C***  THE INITIAL STATE 
C***  ALL = 0, G G = 1, Q Q = 2, G Q = 3, MAJOR=4, MINOR=5

      IFLAVOR = 0

C***  THE CROSS-SECTION WITH CUTS (0)
C***  THE TOTAL CROSS-SECTION WITHOUT CUTS IN A FASTER WAY (1)

      ITOTAL = 0

C***  THE SCALE FOR RENORMALIZATION AND FACTORIZATION
C***  ISCAPT = 0  --> SCALE = MS * SCAFAC 
C***  ISCAPT = 1  --> SCALE = SQRT(MS**2 +PT**2) * SCAFAC
C***                  ONLY FOR ITOTAL = 0 
C***  DEFAULT FOR SCAFAC = 1.0

      ISCAPT = 0
      SCAFAC = 1.D0

C***  THE CUTS ON THE CROSS-SECTION IN PT (DEFAULT: 0,ENERGY )
C***  PT IS ONLY DEFINED FOR POSITIVE VALUES: PTMIN >= 0
C***  EQUAL LOWER AND UPPER CUT GIVES DSIGMA/DPT
C***  ITOTAL = 0 NECESSARY

      PTMIN = 0D0
      PTMAX = ENERGY

C***  THE CUTS ON THE CROSS-SECTION IN Y (DEFAULT: 0,+9.99)
C***  RAPIDITY IS ONLY DEFINED FOR POSITIVE VALUES: YMIN >= 0
C***  EQUAL LOWER AND UPPER CUT GIVES DSIGMA/DY
C***  ITOTAL = 0 NECESSARY
  
      YMIN = 0.D0
      YMAX = +9.99D0

C***  ONLY FOR DISTRIBUTIONS ( ITOTAL = 0 )
C***  DISTINGUISH BETWEEN SQUARKS AND ANTISQUARKS IN THE FINAL STATE
C***  DEFINES THE DIFFERENTIAL CROSS-SECTIONS WITH RESPECT TO
C***  ICHARCONJ =  1  <-- SQUARKS
C***  ICHARCONJ = -1  <-- ANTISQUARKS
C***  ICHARCONJ =  0  <-- AVERAGE OF SQUARKS AND ANTISQUARKS

      ICHARCONJ = 0

C***  IONLYLO = 0 CALCULATES BORN AND NLO CROSS-SECTIONS
C***  IONLYLO = 1 CALCULATES ONLY THE BORN CROSS-SECTION
      
      IONLYLO = 0

C***  THE NUMBER OF VEGAS CALLS ( DEFAULT = 1000 )

      ILO   = 1000
      INLO  = 500

C***  PRINT VEGAS STATISTICS (10) OR NOT (0)

      IPRINT = 0

C --- PRINT THE HEADER ------------------------------------------
      
      CALL PRIHEADSB

C --- INITIALIZE VEGAS ------------------------------------------
     
      CALL RSTART(12,34,56,78)

C ---------------------------------------------------------------
C --- INTEGRATION BY VEGAS --------------------------------------
C --- CALCULATION OF CROSS-SECTIONS AND DISTRIBUTIONS -----------
C --- CAN BE CHANGED --------------------------------------------
C ---------------------------------------------------------------

C***  CALCULATE THE TOTAL CROSS-SECTION WITHOUT CUTS
      ITOTAL = 1

C***  CALCULATE THE CROSS-SECTIONS IN LO AND NLO
      CALL INTEGSB(RESLO,ERRLO,RESNLO,ERRNLO)

C***  PRINT THE CROSS-SECTIONS IN LO AND NLO AND THEIR RELATIVE ERRORS
      CALL PRIRESSB(RESLO,ERRLO,RESNLO,ERRNLO)

      PRINT *

C***  CALCULATE THE DIFFERENTIAL CROSS-SECTION DSIGMA/DPT 
C***  WITH THE SCALE Q**2 = MS**2 + PT**2
C***  FOR PT = 50, 100, 150 GEV
      ITOTAL = 0
      ISCAPT = 1
      DO 100 I = 1,3
         PTMIN = 50.D0 * I
         PTMAX = PTMIN
         CALL INTEGSB(RESLO,ERRLO,RESNLO,ERRNLO)
         CALL PRIRESSB(RESLO,ERRLO,RESNLO,ERRNLO)
 100  CONTINUE

      STOP
      END

C**********************************************************************


\end{verbatim}

\subsection{Control Results}
The sample (main) program of Appendix 4.1 calculates for
squark--antisquark final states:
\begin{itemize}
\item the total cross-section 
\item the differential cross-section $d\sigma/dp_t$, where $p_t$ is
  averaged over squark and antisquark, for $p_t=50,100,150$~GeV and
  the scale $Q=\sqrt{\ms^2+p_t^2}$
\end{itemize}
It produces the following output:
\begin{verbatim}
CROSS-SECTIONS AND DISTRIBUTIONS FOR SQUARK-ANTISQUARK HADROPRODUCTION (IN PB)

PROTON-ANTIPROTON COLLIDER WITH ENERGY =  1800. GEV

IFLAVOR   SCAFAC ISCAPT                              ERR/SIG           ERR/SIG
|   MS    MG   |  | PTMIN PTMAX YMIN YMAX   SIGMA_LO    |     SIGMA_NLO   |
-----------------------------------------------------------------------------
0  280.  200. 1.0 0   0.  1800. 0.00 9.99   0.7028     .0017  0.8333     .0021

0  280.  200. 1.0 1  50.    50. 0.00 9.99   0.2488E-02 .0016  0.3547E-02 .0018
0  280.  200. 1.0 1 100.   100. 0.00 9.99   0.3565E-02 .0015  0.4549E-02 .0018
0  280.  200. 1.0 1 150.   150. 0.00 9.99   0.3102E-02 .0015  0.3683E-02 .0017
\end{verbatim}
This test run takes about 35 minutes on a Silicon Graphics workstation.
It should be noted that the above layout can be freely adapted to the user's
wishes by making appropriate changes in the files \textbf{layout*.f}. 

\subsection{Conventions}

The squared matrix elements (in \textbf{matrix*.f}) are labeled according to a 
fixed set of conventions.
\medspace %

\noindent
The second and third letters are used to represent the type of production 
process:
\begin{itemize}
\item[]  \textbf{SB}: squark--antisquark
\item[]  \textbf{GG}: gluino--gluino
\item[]  \textbf{SG}: squark--gluino
\item[]  \textbf{SS}: squark--squark
\end{itemize}
\medspace

\noindent
The fourth (fifth) letter represents the type of initial state of
parton 1 (parton 2):
\begin{itemize}
\item[]  \textbf{G}: gluon
\item[]  \textbf{Q}: quark
\item[]  \textbf{B}: antiquark
\end{itemize}
For \textbf{QQ} and \textbf{QB} there is an additional flag
$\textbf{IFL}=1/0$ in the parameter list of the function call, indicating 
equal/unequal flavors in the initial state.  
\medspace %

\noindent The sixth letter represents the type of contribution:
\begin{itemize}
  \item[] \textbf{B}: Born
  \item[] \textbf{V}: virtual$\,+\,$soft
  \item[] \textbf{H}: hard
  \item[] \textbf{D}: $\log(\Delta)$
  \item[] \textbf{R}: finite shift in the $\overline{MS}$-scheme to
    restore supersymmetry
  \item[] \textbf{S}: pole part of $1/s_4^2$
  \item[] \textbf{T}: pole part of $1/s_3^2$
  \item[] \textbf{U}: Im[$1/s_4$] Im[$1/s_3$]
  \item[] \textbf{1}: scale dependence of virtual$\,+\,$soft
  \item[] \textbf{2}: scale dependence of $\log(\Delta)$
  \item[] \textbf{3}: scale dependence of hard
\end{itemize}
\medspace

\noindent
e.g.~\textbf{DSBGGV} = squark--antisquark production, gluon--gluon initial 
state, and virtual$\,+\,$soft corrections.

\frenchspacing
 \newcommand{\zp}[3]{{\sl Z. Phys.} {\bf #1} (19#2) #3}
 \newcommand{\np}[3]{{\sl Nucl. Phys.} {\bf #1} (19#2)~#3}
 \newcommand{\pl}[3]{{\sl Phys. Lett.} {\bf #1} (19#2) #3}
 \newcommand{\pr}[3]{{\sl Phys. Rev.} {\bf #1} (19#2) #3}
 \newcommand{\prl}[3]{{\sl Phys. Rev. Lett.} {\bf #1} (19#2) #3}
 \newcommand{\fp}[3]{{\sl Fortschr. Phys.} {\bf #1} (19#2) #3}
 \newcommand{\nc}[3]{{\sl Nuovo Cimento} {\bf #1} (19#2) #3}
 \newcommand{\ijmp}[3]{{\sl Int. J. Mod. Phys.} {\bf #1} (19#2) #3}
 \newcommand{\ptp}[3]{{\sl Prog. Theo. Phys.} {\bf #1} (19#2) #3}
 \newcommand{\sjnp}[3]{{\sl Sov. J. Nucl. Phys.} {\bf #1} (19#2) #3}
 \newcommand{\cpc}[3]{{\sl Comp. Phys. Commun.} {\bf #1} (19#2) #3}
 \newcommand{\mpl}[3]{{\sl Mod. Phys. Lett.} {\bf #1} (19#2) #3}
 \newcommand{\cmp}[3]{{\sl Commun. Math. Phys.} {\bf #1} (19#2) #3}
 \newcommand{\jmp}[3]{{\sl J. Math. Phys.} {\bf #1} (19#2) #3}
 \newcommand{\nim}[3]{{\sl Nucl. Instr. Meth.} {\bf #1} (19#2) #3}
 \newcommand{\el}[3]{{\sl Europhysics Letters} {\bf #1} (19#2) #3}
 \newcommand{\ap}[3]{{\sl Ann. of Phys.} {\bf #1} (19#2) #3}
 \newcommand{\jetp}[3]{{\sl JETP} {\bf #1} (19#2) #3}
 \newcommand{\acpp}[3]{{\sl Acta Physica Polonica} {\bf #1} (19#2) #3}
 \newcommand{\vj}[4]{{\sl #1~}{\bf #2} (19#3) #4}
 \newcommand{\ej}[3]{{\bf #1} (19#2) #3}
 \newcommand{\vjs}[2]{{\sl #1~}{\bf #2}}

\end{document}